\documentclass[aps,prd,amsmath,amssymb]{revtex4-1}
\usepackage{amsmath}
\usepackage{graphicx}
\usepackage{subcaption}
\usepackage{url}

\begin{document}

\title{Static Orbits in Rotating Spacetimes}
\author{Lucas G. Collodel}
 \email{lucas.gardai.collodel@uni-oldenburg.de}
\author{Burkhard Kleihaus}

 \author{Jutta Kunz}

\affiliation{
Institut f\"{u}r Physik, Universit\"{a}t Oldenburg, 
Postfach 2503 D-26111 Oldenburg, Germany}

\begin{abstract}
We show that under certain conditions an axisymmetric 
rotating spacetime contains a ring of points in the equatorial plane,
where a particle at rest with respect to an asymptotic static observer
remains at rest in a static orbit.
We illustrate the emergence of such orbits for boson stars.
Further examples are wormholes, hairy black holes and
Kerr-Newman solutions.
\end{abstract}

\maketitle

\section{Introduction}

Our understanding of the properties of spacetimes largely rests
on the analysis of their geodesics. Therefore the study
of geodesics has been in the focus of gravity research from its
early beginnings. Indeed,
describing the motion of particles and light,
geodesics rapidly established General Relativity, 
predicting the observed orbital precession of Mercury
and the deflection of light by the sun.

In recent years the study of the orbits of stars around
Sgr A$^*$ provided strong evidence for a supermassive
compact central object at the center of the Milky Way,
while observations with the EHT 
should soon provide us with
an image of this object, i.e., presumably the shadow of
a supermassive black hole
\cite{Eckart:2017bhq}.

The orbits around Kerr black holes are well-known
\cite{Chandrasekhar:1985kt}.
In contrast, much less is known about the orbits 
in spacetimes of other compact rotating objects,
such as boson stars, wormholes or hairy black holes.
If such objects were to represent serious contenders
for compact astrophysical objects,
the presence of unique signatures -
not present for Kerr black holes - would be 
highly valuable for their possible identification.

Here we would like to point out such a feature,
which is found in a number of rotating non-Kerr spacetimes.
To this end, let us consider a spacetime with a gravitational potential,
which has a minimum somewhere in a physically accessible region 
(e.g., outside an event horizon or a wormhole throat).
Since the spacetime is rotating, there is frame dragging.
Thus a zero angular momentum particle would be rotating
along with the spacetime, and so would a particle with positive
angular momentum. However, a particle with negative angular
momentum might just stand still with respect to 
an asymptotic static observer, if its angular momentum
would have precisely the right value,
and its radial location would correspond to a minimum
in the potential.

In the following we briefly recall the geodesic equations 
in the equatorial plane for a general rotating metric
and analyze the conditions necessary to have such static orbits,
where massive particles initially at rest
remain at rest for all times.
In particular, we exemplify this mechanism for rotating boson stars,
and point out further examples of spacetimes
with static orbits, before we conclude.

\section{Equatorial geodesics}

Consider an axisymmetric, rotating spacetime,
where the metric coefficients are functions of $r$ and $\theta$. 
In the equatorial plane the line element can be chosen as
\begin{equation}
\label{inte}
ds^2=-Adt^2-2Bdtd\varphi+Ddr^2+Cd\varphi^2,
\end{equation}
where $A$, $B$, $C$ and $D$ are then only functions of the radial
coordinate $r$. 
The action of a test particle moving in this spacetime is given by
\begin{equation}
\label{action}
S=\frac{1}{2}\int g_{\mu\nu}u^\mu u^\nu d\tau,
\end{equation}
which yields the equations of motion in the equatorial plane.
The coordinates $t$ and $\varphi$ are cyclic, 
leading to two first integrals,
identified as the energy per unit mass 
and the angular momentum per unit mass 
of the test particle as measured at spatial infinity, namely
\begin{equation}
\label{cmotion}
E=A\dot{t}+B\dot{\varphi}; \qquad L=-B\dot{t}+C\dot{\varphi}.
\end{equation}

For the equation of motion for the $r$ coordinate
we can employ the four-velocity norm instead, $u^\mu u_\mu=-1$, 
which yields
\begin{equation}
\label{rdotsquared2}
\dot{r}^2=\frac{C}{\Delta D}\left(E-V_+\right)\left(E-V_-\right),
\end{equation}
with the effective potential
\begin{equation}
\label{efpot}
V_{\pm}=\frac{BL\pm\sqrt{\Delta\left(C+L^2\right)}}{C},
\end{equation}
and $\Delta=AC+B^2$. 
Note that in the absence of ergoregions $V_-$ is always negative.
For our purposes it will be sufficient in the following
to only consider $V_+$. 
From eq.~(\ref{cmotion}) we can similarly obtain
\begin{equation}
\label{dvp}
\dot{\varphi}=\frac{B}{\Delta}\left(E-V_\varphi\right),
\end{equation}
where for convenience we introduce an angular potential $V_\varphi=-AL/B$.

\section{Orbits starting from rest}

The Lense-Thirring effect makes it interesting 
to analyze what happens to particles which are initially at rest. 
To accomplish this, the particle must be placed at a point, 
where $E=V_+=V_\varphi$, 
so that both $\dot{r}$ and $\dot{\varphi}$ are zero. 
At this point,
\begin{equation}
\label{condel}
\sqrt{A}\vert_{r_{st}}=E, \quad B\vert_{r_{st}}=-EL.
\end{equation} 

Let us now take a rotating boson star as an example
to illustrate such orbits of massive particles starting from rest.
(See e.g.~\cite{Liebling:2012fv} for a recent
review on boson stars. Orbits in boson star spacetimes
have been considered e.g.~in 
\cite{Eilers:2013lla,Grandclement:2014msa,Meliani:2015zta,Grandclement:2016eng,Grould:2017rzz,Meliani:2017ktw}.)
For the particular example, we choose 
a boson star without self-interaction with
a boson mass $m_b=\sqrt{1.1}$, 
a boson frequency $\omega_s=0.84$, 
and a rotational quantum number $m=1$.

\begin{figure}[h!]
\begin{subfigure}{0.4\textwidth}
\includegraphics[scale=0.58]{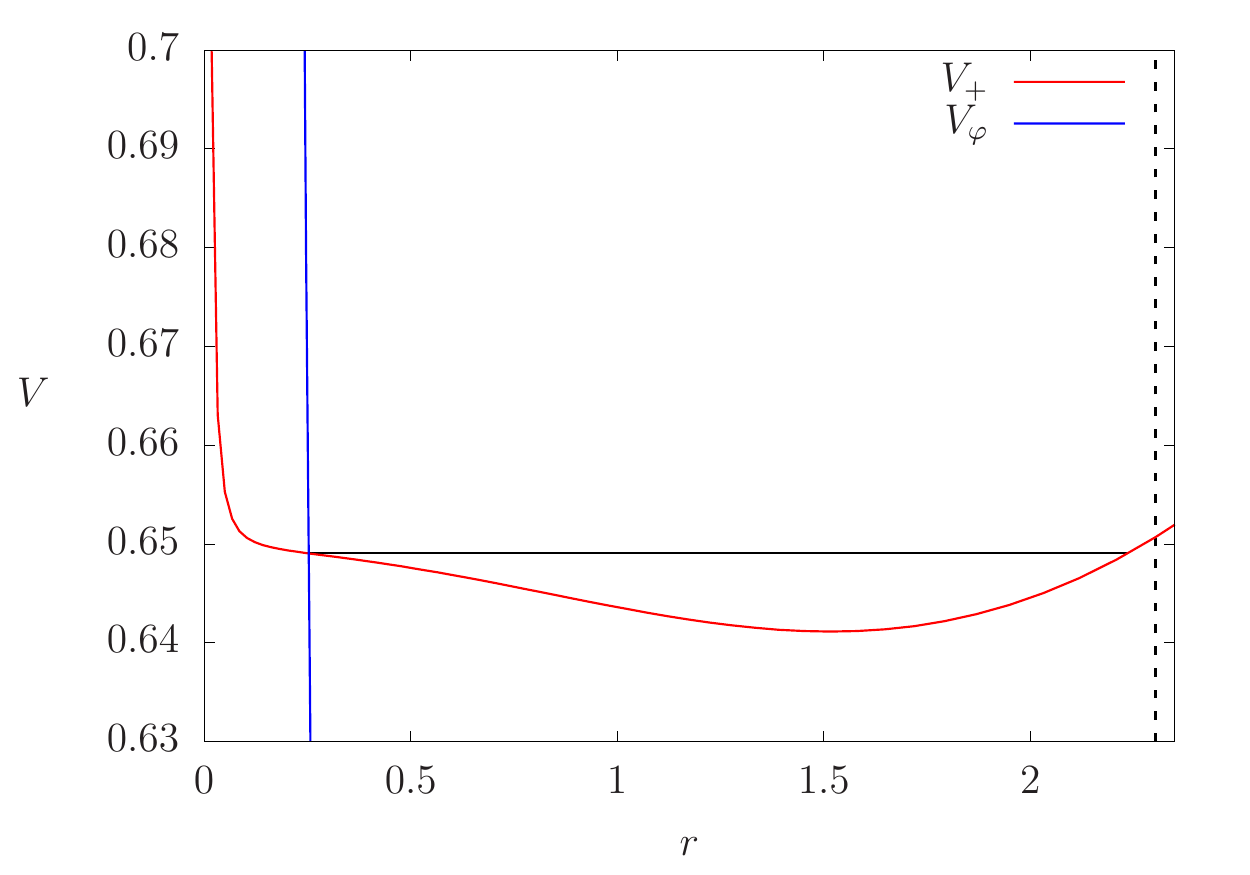}
\label{efpot001}
\end{subfigure} \hspace{0.1\textwidth}
\begin{subfigure}{0.4\textwidth}
\includegraphics[scale=0.58]{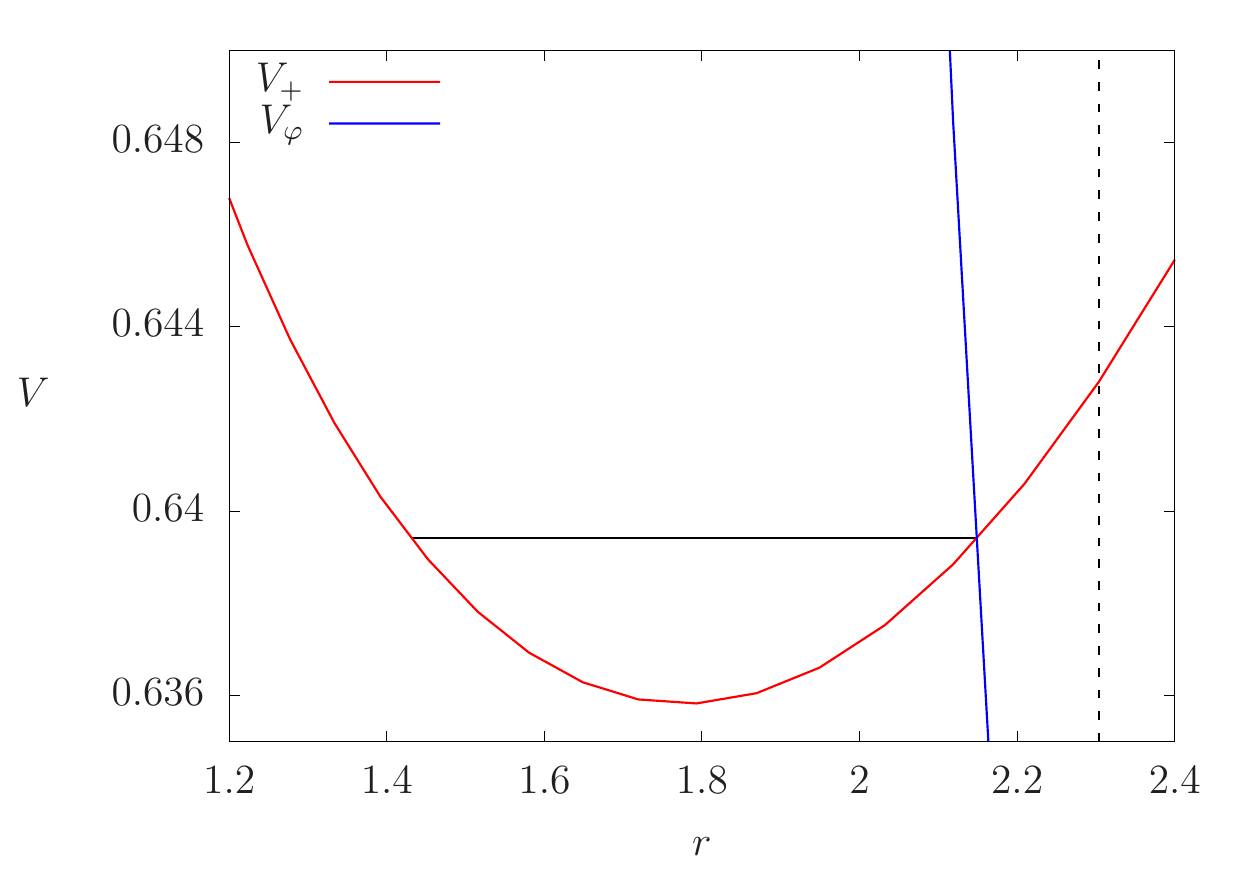}
\label{efpot05}
\end{subfigure}
\caption{Effective potential $V_+$ (red)
and angular potential $V_\varphi$ (blue).
The horizontal solid lines (black) represent the particle energy $E$.
The intersection points with the potentials delimit the radial domain 
of the bound orbit.
The vertical dashed lines (black) indicate the location of the maximum value
of the boson field.
\emph{Left:} Particle with $L=0.01$. 
Its starting rest position is located to the left 
of the minimum of $V_+$.
\emph{Right:} Particle with $L=0.5$. 
Its starting rest position is located to the right 
of the minimum of $V_+$.}
\label{efpotfig}
\end{figure}

Figure \ref{efpotfig} shows the effective potential $V_+$
together with the angular potential $V_\varphi$ for two values 
of the particle angular momentum $L$,
for which eqs.~(\ref{condel}) hold.
Thus the particle energy $E$ is such that the particle starts from rest, 
and the domain of its radial motion corresponds to the
horizontal line. 
When the motion starts from rest for a radial coordinate
$r_{st} < r_{min}$, where $r_{min}$ denotes
the minimum of the potential, 
the particle is pushed \emph{away} from the center 
and engages in the \emph{semi orbit},
co-rotating with the star,
having the largest angular velocity 
(absolute value) at the apocenter 
and being at rest at the pericenter. 
When the motion starts from rest for a radial coordinate
$r_{st} > r_{min}$,
the particle is pulled \emph{towards} the center 
and retains a \emph{pointy petal orbit}, 
counter-rotating with respect to the star 
and having the largest angular velocity (absolute value) 
at the pericenter, while being at rest at the apocenter.
Note, that the occurrence of \emph{semi orbits} and
\emph{pointy petal orbits} was observed in
\cite{Grandclement:2014msa,Grould:2017rzz}.

\section{Orbits remaining at rest}

\begin{figure}[h!]
\hspace*{-0.10\textwidth}
\begin{subfigure}{0.25\textwidth}
\includegraphics[scale=0.7]{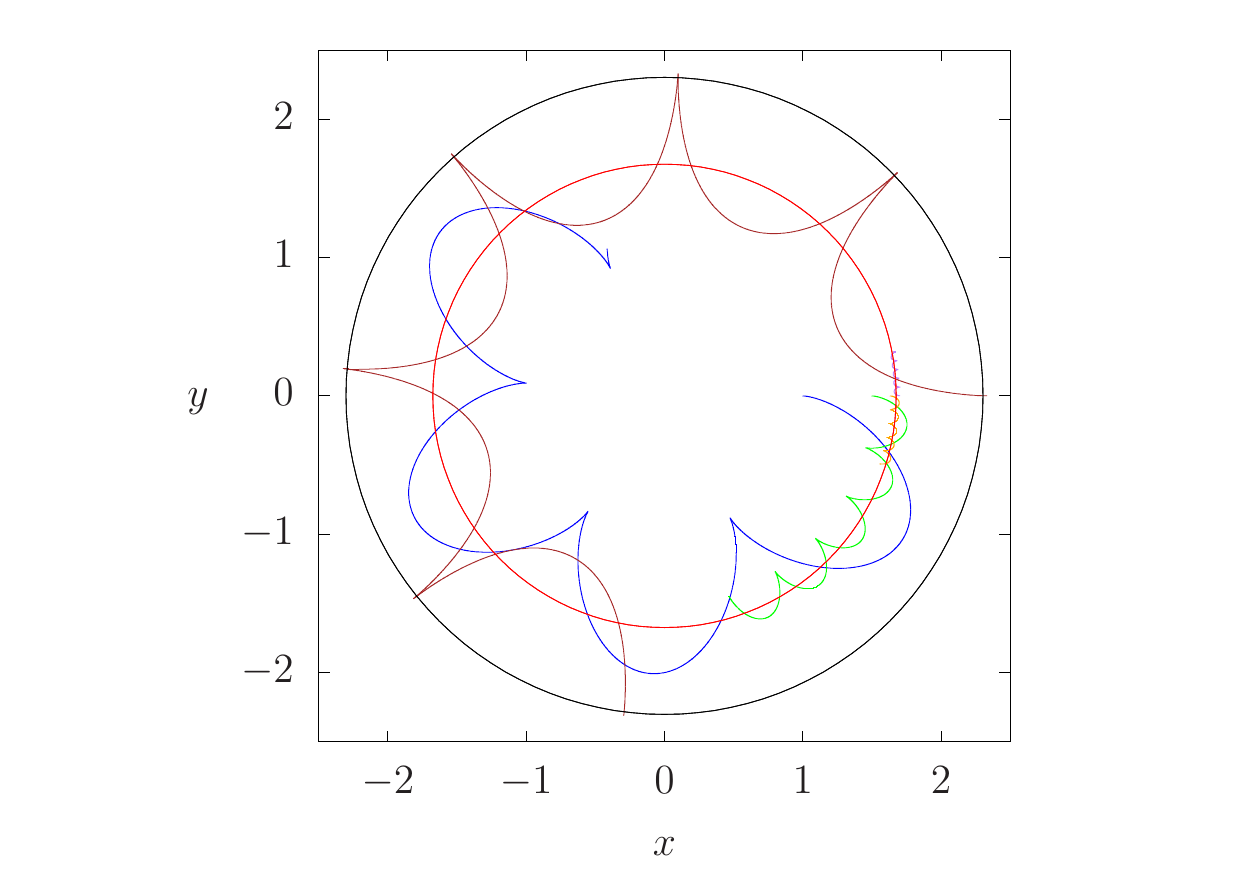}
\label{orb}
\end{subfigure} \hspace{0.20\textwidth}
\begin{subfigure}{0.55\textwidth}
\includegraphics[scale=.85]{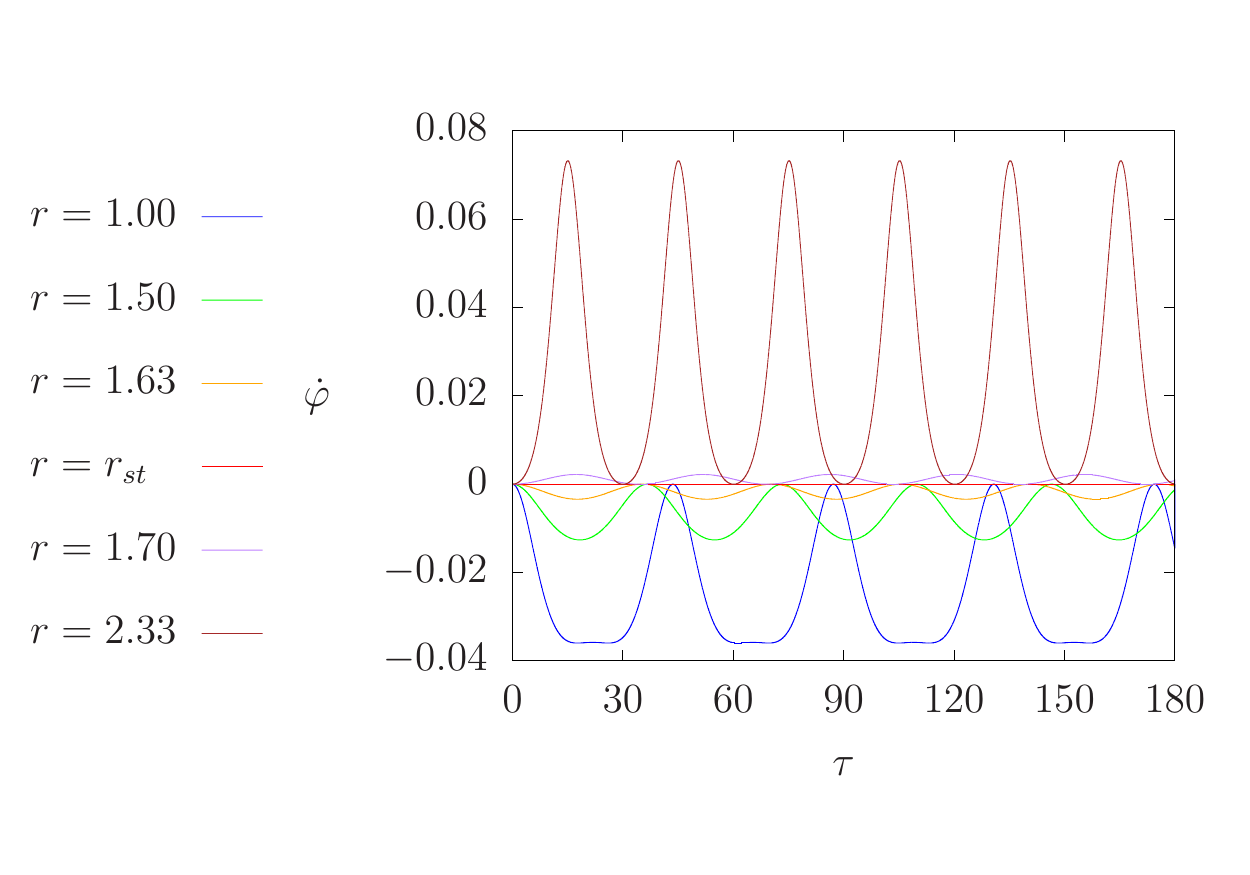}
\label{dphi}
\end{subfigure} 
\caption{Orbits of massive particles released 
from rest at different values of the radial coordinate $r$
(and $\varphi=0$) and evolved for
the proper time interval $\Delta\tau=180$. 
The closer the starting point $r$ is to
the static ring $r_{st}$ (red circle),
the smaller is the amplitude in $\dot{\varphi}$. 
The maximum of the scalar field is also indicated
(black ring).}
\label{orbits}
\end{figure}

We illustrate several orbits starting from rest in Fig.~\ref{orbits}
for the same boson star solution.
These orbits start at different values of the radial coordinate
always from $\varphi=0$, 
and are evolved for the same amount of elapsed proper time
$\Delta \tau$.
Besides the orbits
the oscillating behaviour of $\dot{\varphi}$
is also exhibited (versus the proper time) in Fig.~\ref{orbits}.

As we increase the starting value of $r$, the orbits
change from co-rotating \emph{semi orbits} to counter-rotating
\emph{pointy petal orbits}.
Clearly, in between there arises a critical value of $r$,
$r_{st}=1.6753373$.
At this value a static particle remains static for all times.
Therefore $r_{st}$ represents the static orbit ring.
The occurrence of this static ring is seen from the figure.
The closer the particle starts from the static orbit ring 
the smaller is its angular displacement during the time $\Delta \tau$,
and the smaller is its radial amplitude around the critical point 
$r_{st}=1.6753373$.
Thus in the limit $r \to r_{st}$ the particle remains at rest.

In this case the above condition (\ref{condel})
is met together with the circular orbit condition,
namely $\partial_rV_+=0$. 
But then instead of exhibiting circular motion
the particle will remain at rest in a static orbit
at a distance $r_{st}$ from the center.
Since the relations (\ref{condel}) combined with the circular orbit
condition yield $\partial_rV_+=\partial_rA/(2E)$,
a sufficient and necessary condition for a spacetime
to allow for a stable (unstable) static orbit ring in its equatorial plane
is therefore that its metric component $g_{tt}$ 
presents a maximum (minimum)
anywhere in this plane, $\partial_rA=0$, where $g_{tt}<0$.

Fig.~\ref{orbits} also shows, that
the static orbits (red circle) are located in the star's interior, 
i.e., inside the radius where the scalar field
assumes its maximum value (black circle).
Thus, if a particle is at rest at $r_{st}$ with respect to an
asymptotic static  observer,
it will remain at rest at all times,
provided only the gravitation interaction between this particle
and the boson star is taken into account. 

To guarantee that this feature is independent 
of the chosen parametrization, we need to assure that 
$d\varphi/dr=\dot{\varphi}/\dot{r}=0$ at this point. 
In order to do so, we analyze this ratio at a point 
near the static radius by obtaining the ratio 
of the respective radially perturbed quantities, 
namely $\delta\dot{\varphi}/\delta\dot{r}$. 
We define,
\begin{equation}
V_+=E+\delta V_+; \quad V_{\varphi}=E+\delta V_{\varphi}; \quad A=A_{st}+\delta A; \quad B=B_{st}+\delta B; \quad C=C_{st}+\delta C; \quad D=D_{st}+\delta D.
\end{equation} 
From equations (\ref{rdotsquared2}) and (\ref{dvp}) we see that 
\begin{equation}
\frac{\delta\dot{\varphi}}{\delta\dot{r}}\propto\frac{\delta V_{\varphi}}{\sqrt{\delta V_+}}=\frac{\delta A/E-\delta B E/B}{\sqrt{\delta A/(2E)}}\propto\sqrt{r-r_{st}}.
\end{equation}

Let us now consider circumstances, under which such static rings may occur.
Clearly, stars whose density maximum is located
away from the center represent possible candidates,
with rotating boson stars providing the perfect stage 
for this phenomenon to occur: 
They only interact gravitationally with ordinary matter 
and their energy density distribution is toroidal when rotating. 

\begin{figure}[h!]
\begin{subfigure}{0.3\textwidth}
\includegraphics[scale=0.47]{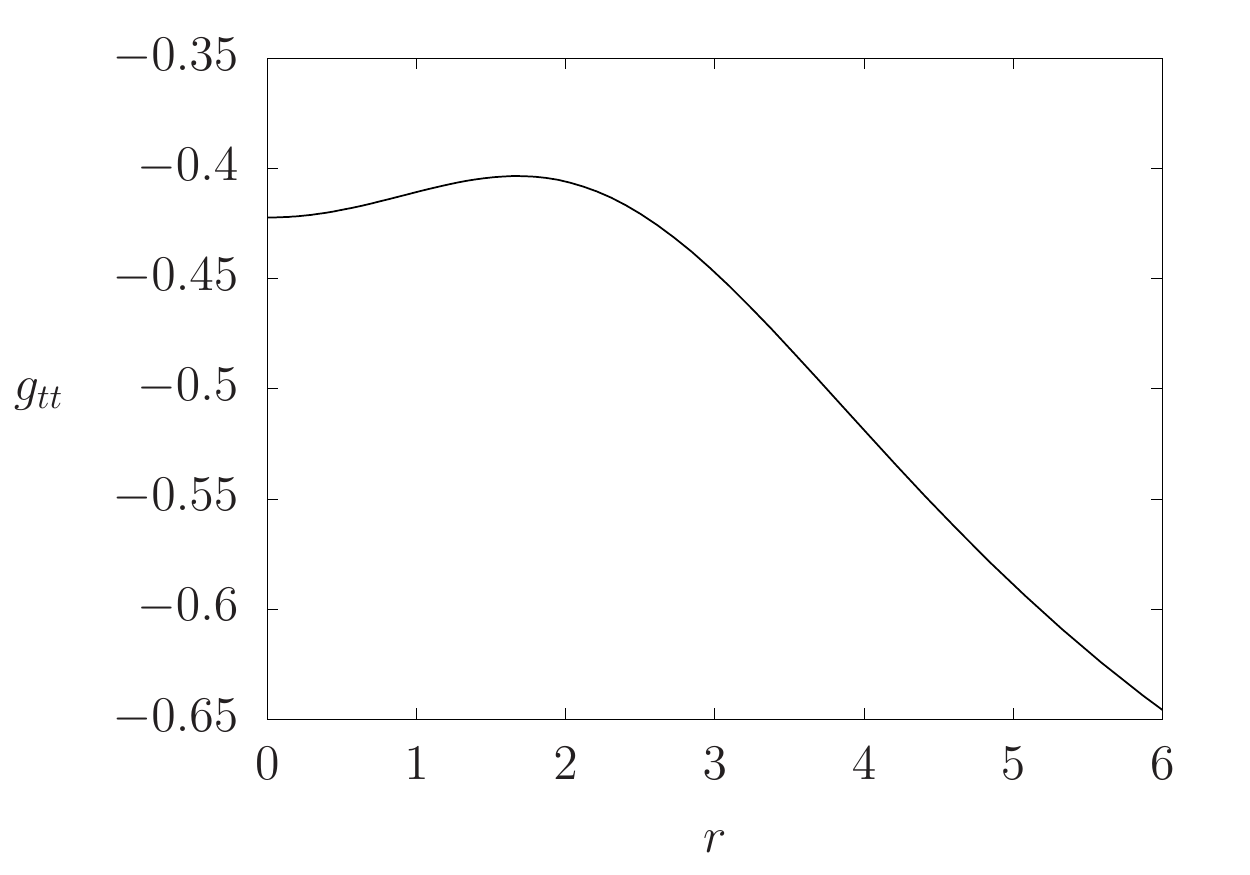}
\label{gttbs}
\end{subfigure} \hspace{0.02\textwidth}
\begin{subfigure}{0.3\textwidth}
\includegraphics[scale=0.47]{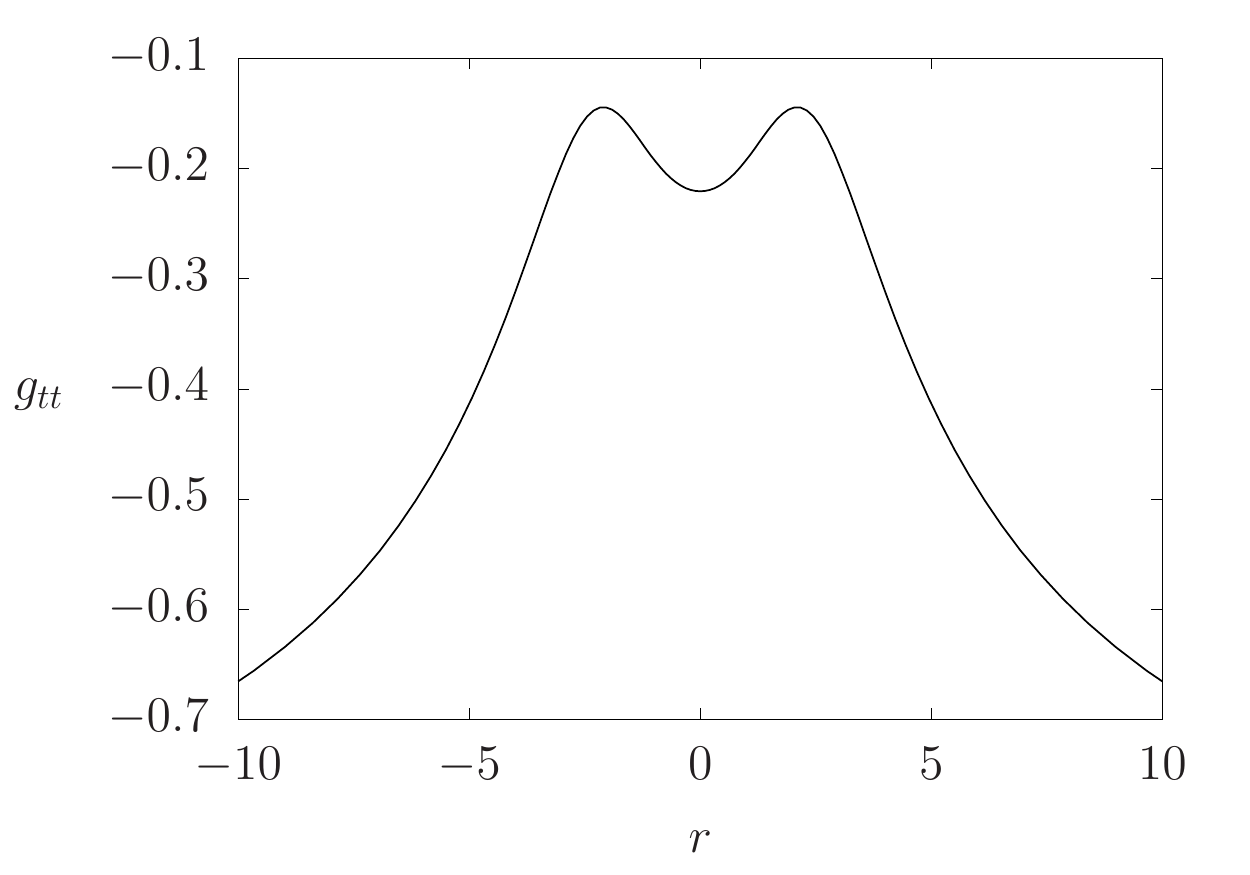}
\label{gttwh}
\end{subfigure} \hspace{0.02\textwidth}
\begin{subfigure}{0.3\textwidth}
\includegraphics[scale=0.47]{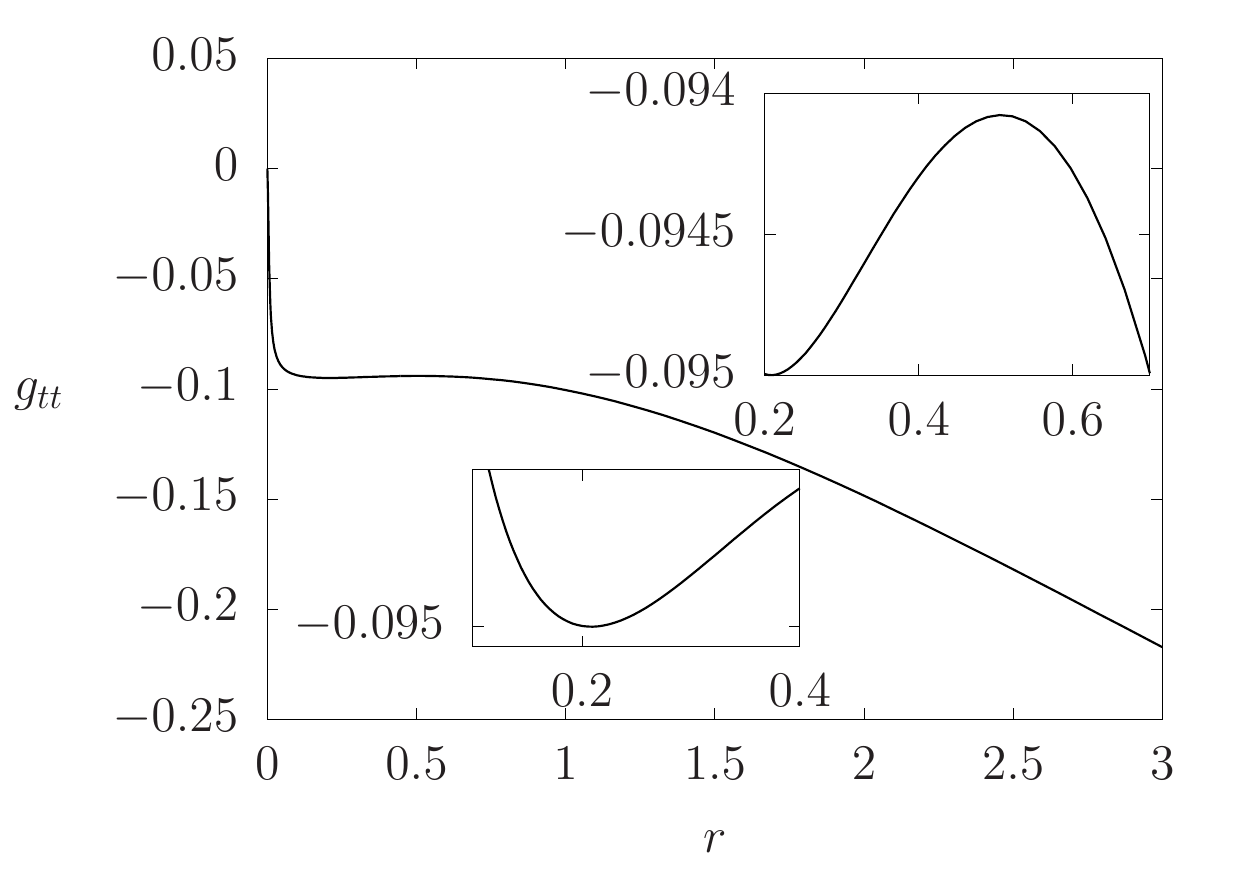}
\label{gttbh}
\end{subfigure} 
\caption{The $g_{tt}$ component for three different systems 
containing bosonic matter. 
\emph{Left:} A pure rotating boson star (without self-interaction). 
\emph{Center:} A traversable rotating wormhole 
immersed in bosonic matter.
\emph{Right:} A rotating black hole with scalar hair.}
\label{gtt}
\end{figure}

To pinpoint the presence of this phenomenon, let us
inspect the $g_{tt}$ component of several rotating spacetimes.
In Fig.~\ref{gtt} we compare the $g_{tt}$ component
of three rotating spacetimes,
all of which contain a complex scalar field: a boson star, 
a traversable wormhole immersed in bosonic matter and a hairy black hole. 
The boson star features a local maximum of $g_{tt}$, 
corresponding to a local minimum of the effective potential $V_+$,
where a test particle might be trapped at rest. 
The solution depicted corresponds to the one analyzed above.
However, this feature is rather generic for boson star solutions,
arising also for quartic and sixtic self-interaction potentials
\cite{Kleihaus:2005me,Kleihaus:2007vk,Kleihaus:2011sx}
as well as for rotationally and radially excited boson stars
\cite{Collodel:2017biu}.

The rotating wormhole solution contains a phantom field
besides the complex scalar field, which in the
example depicted is non-interacting
\cite{Hoffmann:2017}.
Such solutions represent rotating generalizations
of the wormhole solutions studied in 
\cite{Dzhunushaliev:2014bya}.
The metric function $g_{tt}$ exhibited in Fig.~\ref{gtt}
contains three local extrema,
two symmetric maxima and a minimum at the throat (at $r=0$). 
It is thus possible to have particles at (unstable)
rest exactly at the throat in between the two universes 
and also at (stable) rest at a certain position $\pm r_{st}$ 
in each of the universes.

Rotating black holes carrying complex scalar field hair
arise for a non-interacting boson field as well as in the
case of self-interaction
\cite{Herdeiro:2014goa,Kleihaus:2015iea,Herdeiro:2015tia}.
As seen in the figure, for a sufficiently small horizon radius,
the metric function $g_{tt}$ features both a local maximum and a local 
minimum, possessing thus two static orbit rings, one stable and one
unstable, respectively.
The hairy black hole depicted is 
based on a quartic potential \cite{Kleihaus:2015iea}. 
Note, that the radial coordinate in the figure is shifted
such that the horizon radius is located $r_H=0$.

The static orbit ring is, however, not an exclusive feature 
of spacetimes warped by a spin-0 field. 
The $g_{tt}$ component of the well known Kerr-Newman solution 
also contains a local maximum at $r_{st}=Q^2/M$. Not only is this radius
independent of the angular momentum but is also present at the same 
location in the non-rotating Reissner-Nordstr\"{o}m solution.
Surprisingly, the authors could not find any reference to these findings
in the literature. We emphasize that the equations of motion are 
geodesics, namely for test particles without electric charge. 
The  close relation between $g_{tt}$ and the energy per unit mass of the
orbiting particle, eq. (\ref{condel}), demands a more careful analysis
for these spacetimes. The only possible observational scenario is that
of a naked singularity, where the static ring is always present, whether
there is rotation or not. 
However, when the solution is a black hole, 
the static ring is always hidden behind the event horizon. 
Static rings are located behind both horizons 
if $J^2\geq\alpha^2-\alpha^4$, where $J$ is the black hole's 
angular momentum and $\alpha=Q/M$.
By writing $J^2=\alpha^2(1-\alpha^2)+\delta$, with
$\delta\geq 0$, we obtain $\alpha^2\leq 1-\sqrt{\delta}$ as a further 
constraint on the charge.  The 
static ring is also found in the presence of charged black holes in higher
dimensions. The five dimensional Breckenridge-Myers-Peet-Vafa (BMPV) 
spacetime (for details on the geodesics see 
\cite{Gibbons:1999uv,Herdeiro:2000ap,Diemer:2013fza})
possesses two orthogonal rings (parametrized by $\varphi$ and $\psi$) 
at $r_{st}=\sqrt{\mu}$, where $\mu$ is again a parameter that depends 
only on the mass and charge.

\section{Conclusions}

We have shown that under specific conditions 
a rotating spacetime may possess a ring in the equatorial plane,
where massive particles initially at rest 
with respect to an asymptotic static observer remain at rest. 
It is worth mentioning that this phenomenon 
of static orbits is a purely inertial one.
(This is in contrast to the equilibrium positions
found for charged particles orbiting charged black holes.)

The physical conditions to be met in order to allow for such static
orbits require, that besides a precise balance between
the angular momentum and the frame dragging,
the gravitational potential contained in the
metric component $g_{tt}$ should possess a minimum,
such that a particle sitting
at the corresponding radius $r_{st}$ is neither pulled towards the center
nor pushed away from it. Clearly, the extremum of $g_{tt}$
and the extremum of the effective potential must therefore coincide
at $r_{st}$.

Let us note that somewhat
similar conditions are found for null geodesics \cite{Grandclement:2016eng}.
Here, the presence of so-called static {\sl light points},
requires furthermore that their spatial location
coincides with the onset or termination of an ergoregion, 
$A\vert_{r_{st}}=0$, and that the photons have zero energy. 

While we have exemplified our analysis in detail for
boson stars, we have shown, 
that static orbits arise not only for boson stars 
but also for Ellis wormholes 
immersed in rotating bosonic matter,
as well as for hairy black holes
or Kerr-Newman solutions.

While one might argue that it might be unlikely 
to expect nature to produce such a fine tuned scenario, 
we have verified that geodesics 
followed by particles initially at rest near the (stable) static radius 
are characterized by slow motion. The closer they start from $r_{st}$ 
the smaller are their maximal radial and angular velocities. 
This opens the possibility of observing this phenomenon, 
if rotating compact objects such as the above
examples would indeed exist in nature. 

Therefore, observing static, or quasi-static, astrophysical objects
in a region which otherwise indicates the presence
of a strong gravitational field
could offer support for the presence of
a type of compact object that differs from Kerr black holes.

\section{Acknowledgments}

We would like to acknowledge support by the DFG Research Training Group 1620
{\sl Models of Gravity} as well as by FP7, Marie Curie Actions, People,
International Research Staff Exchange Scheme (IRSES-606096), 
COST Action CA16104 {\sl GWverse}.
BK gratefully acknowledges support 
from Fundamental Research in Natural Sciences
by the Ministry of Education and Science of Kazakhstan.

\end{document}